%% file: main.tex
\documentclass[twocolumn,trackchanges]{aastex631}
\pdfoutput=1 
\usepackage[T1]{fontenc}
\usepackage[figure,figure*]{hypcap}
\usepackage{amsmath,amstext,bm}
\usepackage{graphicx,textcomp,fancyhdr,hyperref}

\shorttitle{KELT-18~\textnormal{b} has a Polar Orbit}
\shortauthors{Rubenzahl et al.}

\DeclareRobustCommand{\okina}{%
\raisebox{\dimexpr\fontcharht\font`A-\height}{%
    \scalebox{0.8}{`}%
  }%
}
\newcommand{\Porb}{P_\text{orb}}
\newcommand{\istar}{i_\star}
\newcommand{\iorb}{i_\text{orb}}

\newcommand{\veq}{v_{eq}}
\newcommand{\sini}{\sin i_\star}
\newcommand{\vsini}{\veq\sini}

\newcommand{\avgmu}{\langle\mu\rangle}
\newcommand{\Teff}{T_\text{eff}}
\newcommand{\Msun}{M_\odot}

\newcommand{\Rjup}{R_J}

\newcommand{\ms}{m\,s\ensuremath{^{-1}}}
\newcommand{\kms}{km\,s\ensuremath{^{-1}}}

\begin{document}

\title{KPF Confirms a Polar Orbit for KELT-18~b}

\input{authors}

\newcommand{\lambest}{-94.8}
\newcommand{\elambestlo}{0.7}
\newcommand{\elambesthi}{0.7}
\newcommand{\bestlam}{\lambest \pm \elambestlo{}^\circ}
\newcommand{\psibest}{91.7}
\newcommand{\epsibestlo}{1.8}
\newcommand{\epsibesthi}{2.2}
\newcommand{\bestpsi}{\psibest_{-\epsibestlo}^{+\epsibesthi}{}^\circ}

\begin{abstract}

We present the first spectroscopic transit results from the newly commissioned Keck Planet Finder on the Keck-I telescope at W. M. Keck Observatory. We observed a transit of KELT-18~b, an inflated ultra-hot Jupiter orbiting a hot star ($\Teff = 6670$~K) with a binary stellar companion. By modeling the perturbation to the measured cross correlation functions using the Reloaded Rossiter-McLaughlin technique, we derived a sky projected obliquity of $\lambda = \bestlam$ ($\psi = 93.8_{-1.8}^{+1.6}{}^\circ$ for isotropic $\istar$). The data are consistent with an extreme stellar differential rotation ($\alpha = 0.9$), though a more likely explanation is moderate center-to-limb variations of the emergent stellar spectrum. We see additional evidence for the latter from line widths increasing towards the limb. Using loose constraints on the stellar rotation period from observed variability in the available TESS photometry, we were able to constrain the stellar inclination and thus the true 3D stellar obliquity to $\psi = \bestpsi$. KELT-18~b could have obtained its polar orbit through high-eccentricity migration initiated by Kozai-Lidov oscillations induced by the binary stellar companion KELT-18~B, as the two likely have a large mutual inclination as evidenced by \textit{Gaia} astrometry. KELT-18~b adds another data point to the growing population of close-in polar planets, particularly around hot stars.
\end{abstract}

\section{Introduction} \label{sec5:intro}

KELT-18~b is an ultra-hot Jupiter discovered by the KELT transit survey~\citep{McLeod2017}. The $1.57~\Rjup$ planet orbits its F5 type ($6670$~K) host star every 2.87~days. Hot stars ($\gtrsim 6250$~K) with hot Jupiters (HJs) have been observed to have a broad range of obliquities, where the obliquity is defined as the angle between the host star's rotation axis and the planet's orbital plane. Conversely, HJs orbiting cooler stars ($\lesssim 6250$~K) tend to be aligned with their host star's rotation axis \citep{Winn2010hotStars, Schlaufman2010, AlbrechtReview}. The transition temperature is near the Kraft Break \citep{KraftBreak}, suggesting realignment mechanisms that are effective for cooler stars--which have convective envelopes and strong magnetic fields--but are ineffective for hotter stars--which have radiative envelopes and weak magnetic fields \citep{Albrecht2012, Dawson2014}. 

While it is still an unsolved problem, the origins of HJs are likely a combination of multiple formation channels (see \citealt{DawsonJohnsonReview} for a review), namely in-situ formation, disk migration, and high eccentricity migration (HEM). HEM likely plays a significant role in shaping the overall HJ population \citep{Rice2022} but must be triggered by an additional body in the system. This could be another planet, in the case of planet-planet scattering \citep{Rasio1996} or von-Zeipel-Kozai-Lidov\footnote{See \citet{ZKL} for a historical monograph.} oscillations \citep[ZKL;][]{vonZeipel1910, Kozai1962, Lidov1962} induced by an outer planetary \citep{Naoz2011, Teyssandier2013} or stellar companion \citep{Fabrycky2007}. In the HEM scenario, the HJ originally formed beyond the water ice-line ($\sim$2~AU) where giant planet formation is efficient \citep{Pollack1996}. The orbital eccentricity was increased through interactions with a perturbing companion until the planet's periastron distance became small enough for tides to dissipate energy and transfer orbital angular momentum to the star, causing the orbit to shrink and circularize. Either this HEM process, or perhaps a primordial misalignment of the protoplanetary disk \citep{Batygin2012}, leaves the HJ on an orbit that may be tilted by a large angle relative to the stellar equatorial plane. Only the HJs around stars cooler than the Kraft Break were then able to realign their host star's rotation axis.

The spin-orbit angle is usually measured as projected on the plane of the sky ($\lambda$), but for systems in which the inclination of the host star's rotation axis ($\istar$) can be inferred, the true 3D obliquity can be derived ($\psi$). Recently, \citet{Albrecht2022} noted that hot Jupiters around hot stars do not span the full range of $\psi$, but instead show a preference for near polar orbits ($80^\circ$--$120^\circ$). However, there are still too few systems to be sure that the obliquity distribution has a peak near $90^\circ$  \citep{Siegel2023, Dong2023}. If there is a ``polar peak'' it would have important theoretical implications on plausible HJ formation mechanisms, which predict different obliquity distributions \citep[see][]{AlbrechtReview}. For small planets with massive outer companions, secular resonance crossing in the disk dispersal stage may produce polar orbits \citep{Petrovich2020}. For giant planets, an initially inclined orbit inherited from a torqued protoplanetary disk in the presence of a binary companion can give the necessary starting point for subsequent ZKL-driven migration to create a polar HJ \citep{Vick2023}.

The most commonly employed method for measuring the projected obliquity of a star with a transiting planet is to obtain high-resolution spectra throughout a transit and model the Rossiter-McLaughlin effect \citep{Rossiter1924, McLaughlin1924},
which results from
the planet's obscuration of part of the rotating stellar photosphere.
This effect is often
modeled as an anomalous radial velocity (RV) signal, but measuring precise RVs for hot stars can often be challenging due to their fast rotation rates. The projected equatorial rotation velocity, $\vsini$, broadens (and blends) spectral lines, diminishing the Doppler information content \citep{Bouchy2001}. As a result, stars with $\vsini \gtrsim 10$~{\kms} are usually not amenable to anomalous-RV modeling. However, these fast rotating stars lend themselves to more detailed and direct methods of measuring the stellar obliquity. 
The Reloaded RM (RRM) method, developed by \citet{Cegla2016}, directly models the distortion of the line profile by the transiting planet. By subtracting an out-of-transit reference CCF (representing the star alone) from each in-transit CCF (corresponding to the stellar line profile integrated over the full disk, minus the integrated line profile from within the patch of the star beneath the planet's shadow), the resulting signal represents the ``local'' CCF, i.e., the spectrum originating from the portion of the star obscured by the transiting exoplanet (CCF$_\text{loc}$). The RRM method is also sensitive to stellar differential rotation, should the planet be highly misaligned so that it transits a wide range of stellar latitudes \citep{RoguetKern2022}.

In this paper we report our derivation of the obliquity of the host star in the KELT-18 system based on a time series of spectra taken during a transit of KELT-18~b with the Keck Planet Finder (KPF). By modeling the spectra according to the RRM method, we found the orbit of KELT-18~b to be nearly perpendicular to the star's equatorial plane. In Section~\ref{sec5:systemprops} we derive stellar properties, reexamine the rotation period with \textit{TESS} photometry, and identify the nearby star KELT-18~B as a bound companion. We describe the Keck Planet Finder and our observations in Section~\ref{sec5:observations}, the RRM modeling procedure in Section~\ref{sec5:obliquity}, and dynamical implications for the KELT-18 system in Section~\ref{sec5:dynamics}.

\section{KELT-18 System}\label{sec5:systemprops}

\input{kelt18systemprops}

KELT-18 is a rapidly rotating F4 V star with one known transiting exoplanet, discovered by \citet{McLeod2017} (hereafter M17), and a stellar neighbor. M17 derived robust stellar properties using high resolution spectra, SED fitting, photometry, and evolutionary modeling in a global fit with their transit model. We adopted their best-fit stellar and transit parameters for our analyses herein, with two distinctions noted below. Table~\ref{tab5:systemprops} lists the full set of adopted parameters.

For the transit midpoint and orbital period of KELT-18~b, we adopted the improved ephemeris of \citet{Ivshina2022} which implies an uncertainty of only 20~sec in the predicted transit midpoint on the night of our spectroscopic observations (Section~\ref{sec5:observations}).

For the projected stellar rotation velocity $\vsini$, M17 noted that their value of $12.3 \pm 0.3$~{\kms} obtained from a TRES \citep{TRES} spectrum is likely an overestimate as the method they used conflates macroturbulence and rotation. M17 also measured a value of $10 \pm 1$~{\kms} using a HIRES \citep{HIRES} spectrum and the \texttt{SpecMatch-Synthetic}~\citep{Petigura2015} framework, but did not adopt this value because the best-fit $\Teff$ was outside the range 4800--6500~K over which the code had been calibrated. Since then, the \texttt{SpecMatch-Emperical}~\citep{smemp} tool was developed to derive stellar properties for a wider range of effective temperatures (3000--7000~K) by interpolating a grid of library spectra obtained with Keck/HIRES. We ran the same HIRES spectrum obtained by M17 through \texttt{SpecMatch-Emperical} to obtain new estimates of $\Teff$, Fe/H, and $R_\star$. The resulting $\Teff = 6330 \pm 110$~K is cooler than the $6670\pm 120$~K value of M17 and is within the \texttt{SpecMatch-Synthetic} regime. We therefore ran \texttt{SpecMatch-Synthetic} on the same HIRES spectrum and found $\Teff = 6530 \pm 100$~K and $\vsini = 10.4 \pm 1.0$~{\kms}. Thus, the true value of $\vsini$ is likely in the 9--12~{\kms} range. All this together informs our adoption of an informed prior on $\vsini$ of $10.4 \pm 1$~{\kms} for our spectroscopic transit analysis in Section~\ref{sec5:obliquity}.




\subsection{Stellar rotation period}\label{sec5:rotation-period}

A significant peak at 0.707~days was observed in the Lomb-Scargle periodogram of the KELT photometry, which M17 interpreted as the rotation period of KELT-18. Given the measured stellar radius, this implied an equatorial rotational velocity of $\veq = 134$~{\kms}. While large, it is not atypical for stars of KELT-18's $\Teff$ and $\log g$ to have rotation speeds on the order of $100$~{\kms}. Combining this with their measured $\vsini$ of $12.3$~{\kms}, \citealt{McLeod2017} noted that the star must have an inclination of $\sim$$5^\circ$. In other words, we are observing KELT-18 nearly pole-on. Since the planet's orbit is viewed at high inclination, the implication was that the planet's orbit is nearly polar.

\begin{figure}
    \centering
    \includegraphics[width=0.495\textwidth]{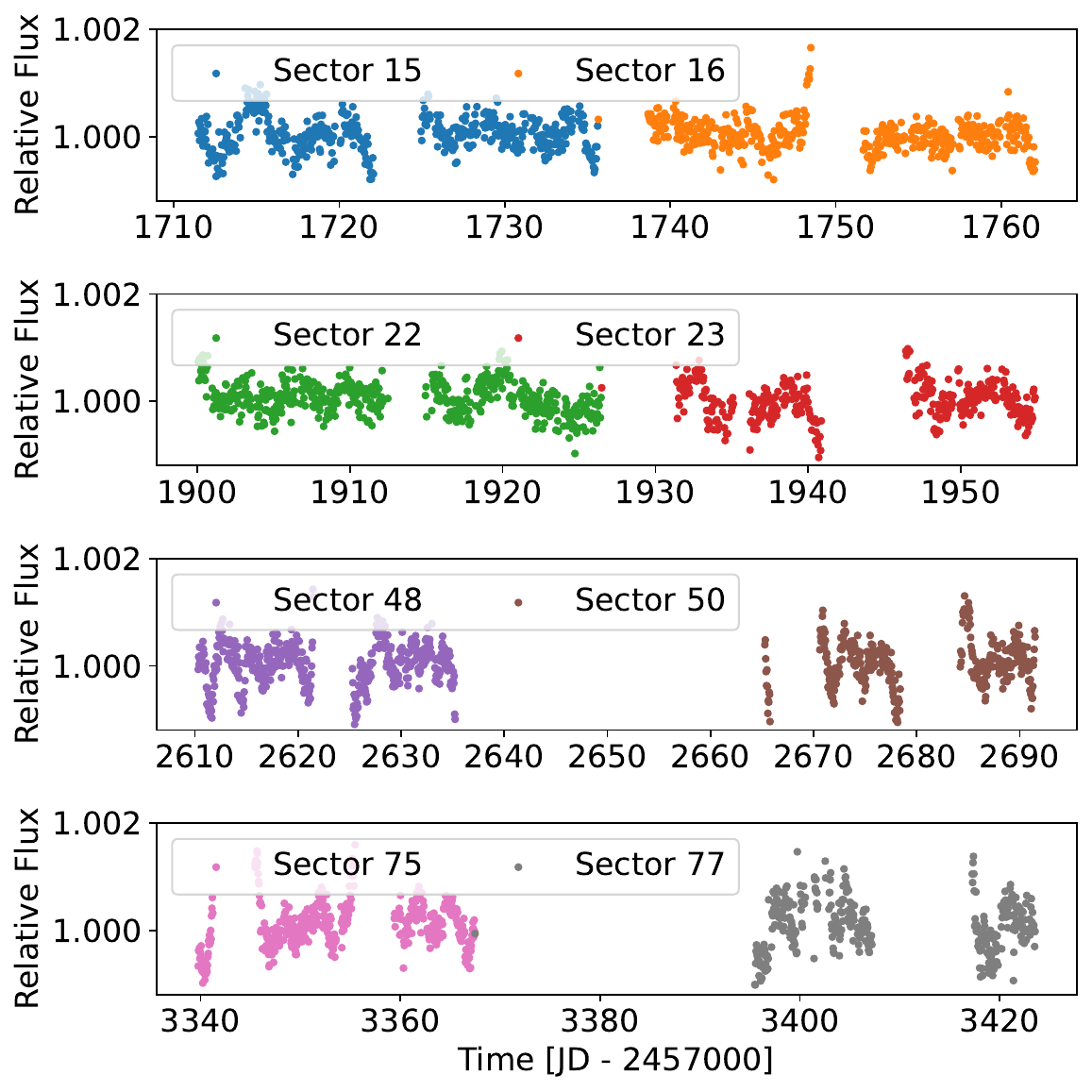}
    \includegraphics[width=0.495\textwidth]{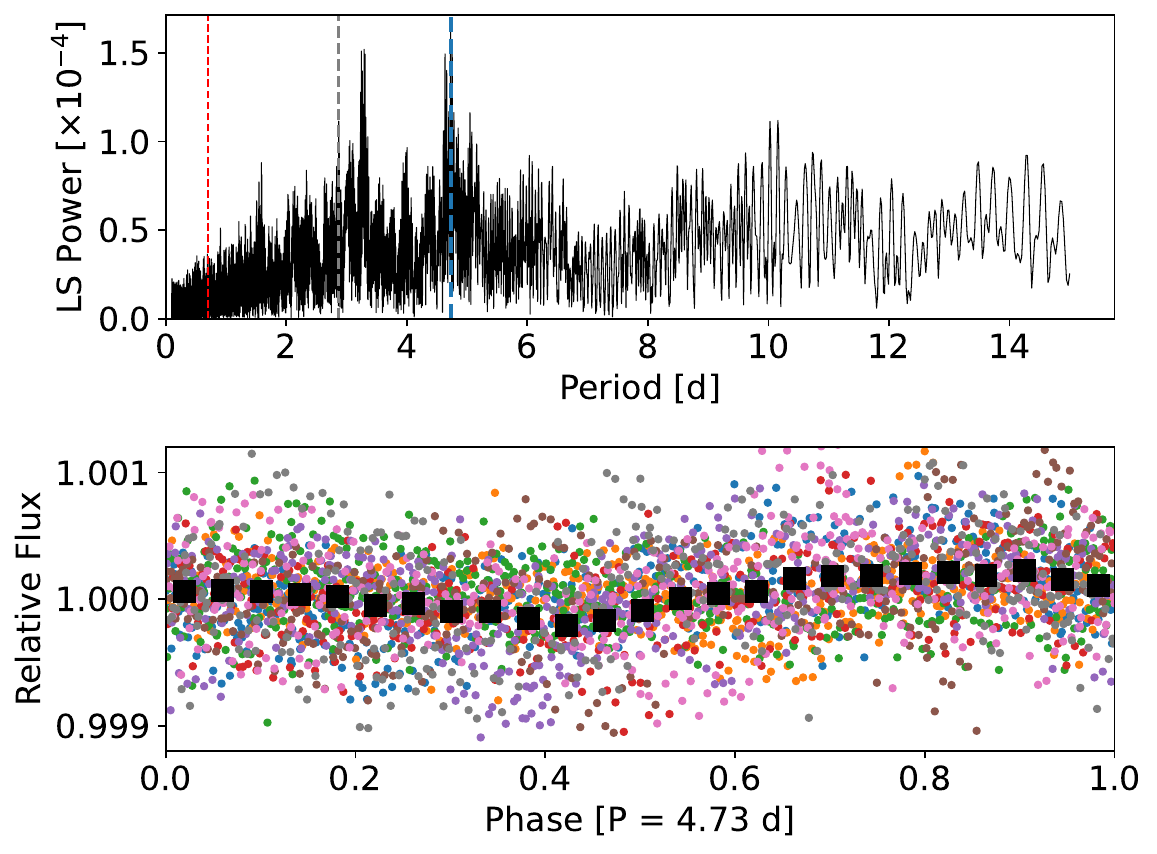}
    \caption{Analysis of the rotation period using TESS photometry. The top three panels show the 1~hour binned TESS photometry described in Section~\ref{sec5:rotation-period}. The fourth panel displays a Lomb-Scargle periodogram. We marked the 0.707~d period from M17 with a red dashed line, KELT-18's orbital period with a grey dashed line, and the maximum peak with a blue dashed line. The bottom panel shows all TESS data points (same color-coding per sector as above) phase-folded to the period with maximum power. Black squares show the phased data evenly binned with bin size 0.05.}
    \label{fig5:lightcurve}
\end{figure}

Since the rotation signal in the KELT photometry appeared small compared to the measurement noise, we downloaded the available TESS photometry for KELT-18 to search for variability. KELT-18 was observed by TESS as TIC 293687315 (TOI-1300) in sectors 15, 16, 22, 23, 48, 50, 75, and 77. We downloaded the 1800~s cadence data for sectors 48 and 50, 120~s data for sector 75 and 77, and the 600~s data for the other sectors. We selected data processed by the TESS Science Processing Operations Center pipeline \citep{TESSSPOC}, removed flagged values, stitched the six sectors together, and binned to a common 1~hour cadence using \texttt{lightkurve} \citep{lightkurve}. The resulting light curve is shown in Figure~\ref{fig5:lightcurve} along with its Lomb-Scargle periodogram.

The TESS periodogram does not contain any significant peaks below $\sim$3~days. There is a clustering of peaks around $\sim$5~days with a maximum power at 4.76~d, and variability on this timescale is visible by eye in the TESS photometry. Stars with KELT-18's $\Teff$ tend to have rotation rates $< 8$~d at $>3\sigma$ \citep{Bouma2023}, so a $\sim$5~day rotation period would be reasonable. However, further inspection at the per-sector level (see Appendix~\ref{sec:rot-appendix}) reveals that this $\sim$5 day periodicity is only dominant in sectors 15, 23, and 48. Other sectors show muted variability or strong peaks at other periods. Interactions between multiple spot groups, and/or the presence of differential rotation, can yield multiple periodogram peaks which may further wander in time \citep{Blunt2023}. Altogether, a precise and robust value for the stellar rotation period is not well-determined by the TESS photometry, though it does bring the previous value of 0.707~days into question.  

A rotation period of 4.76~days would imply an equatorial rotation speed of $\veq \sim 20$~{\kms}. Given our spectroscopic measurement of $\vsini$ is only $10.4\pm 1$~{\kms}, it remains likely that KELT-18 is viewed at low inclination. Conservatively, we adopt a uniform $< 8$~d prior on the rotation period based on the $3\sigma$ empirical boundary defined by \citealt{Bouma2023} and the TESS photometry showing variability on the order of several days. Using the methodology of \citep{MasudaWinn2020}, this loose constraint translates into a stellar inclination of $\istar = 31^{+35}_{-15}{}^\circ$.

\subsection{KELT-18~B: Neighbor or Companion?}\label{sec5:companion}

M17 noted a stellar neighbor at 3''.43 separation from KELT-18~B. The neighbor is fainter at $K = 12.9 \pm 0.2$. Under the assumption that the neighbor is at the same distance as KELT-18, M17 obtained ${\Teff}_B \sim 3900$~K. The available astrometry were not precise enough to identify the neighbor as comoving based on its proper motion, though the relatively small sky density of stars in the field around KELT-18 (at high galactic latitude) makes a chance alignment at such a small angular separation unlikely ($>$3$\sigma$).

KELT-18 and KELT-18~B appear in the catalog of stellar companions to TESS Objects of Interest of \citet{Behmard2022} (B22). B22 applied the methodology of \citet{Oh2017} to the \textit{Gaia} DR3 astrometry \citep{gaiadr3} to determine the likelihood the two stars are comoving. The method propagates the astrometric uncertainties from \textit{Gaia} into a likelihood ratio comparing the comoving hypothesis ($\mathcal{L}_1$) to the null (not comoving) hypothesis ($\mathcal{L}_2$). B22 added a jitter term to account for any unknown systematic effects to improve the reliability of this hypothesis testing. B22 computed $\ln(\mathcal{L}_1/\mathcal{L}_2) =  4.83$ for KELT-18 and KELT-18~B, giving strong evidence for the comoving hypothesis. They computed a stellar mass for KELT-18~B of $0.575_{-0.026}^{+0.025}~\Msun$ using \texttt{isoclassify} \citep{isoclassify} and the \textit{Gaia} magnitudes, in agreement with the ${\Teff}_B$ estimate from M17. Given the consistent parallaxes ($3.16\pm0.01$~mas for the primary and $3.23\pm0.04$~mas for the secondary), B22 computed a binary separation from the \textit{Gaia} astrometry of 1082~AU. The relative velocity vector in the sky plane is thus $0.695 \pm 0.075$~{\kms}. For reference, if both stars were orbiting in the sky-plane on circular orbits, their relative velocity would be 1.8~{\kms}. So, unless the two stars have a significant relative radial velocity (which \textit{Gaia} did not measure), they are likely bound.

\section{Observations}\label{sec5:observations}

\begin{figure*}
    \centering
    \includegraphics[width=0.9\textwidth]{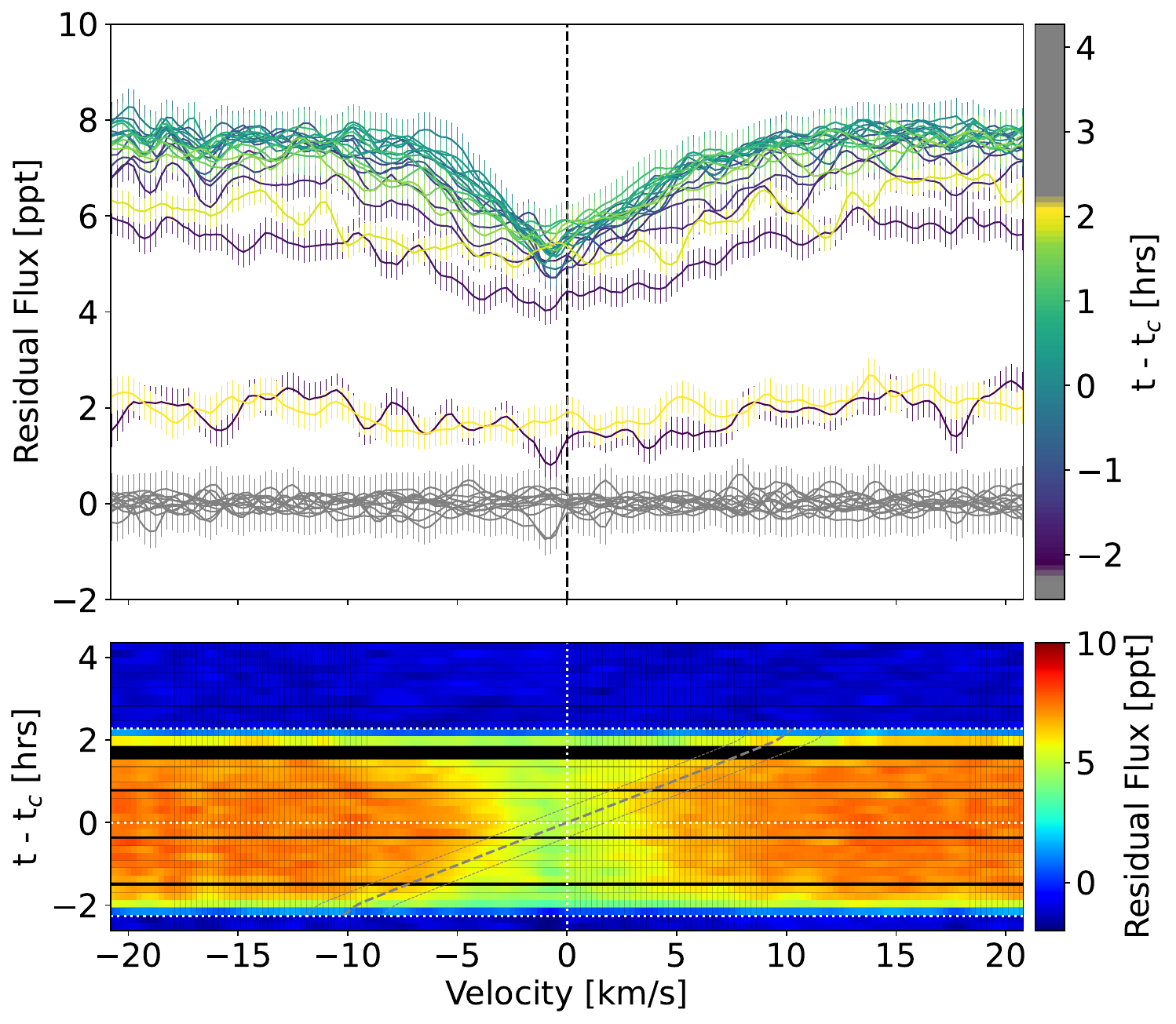}
    \caption{The timeseries of CCF$_\text{loc}$ measured with KPF. The top panel shows each 1D CCF$_\text{loc}$, with out-of-transit observations colored grey and in-transit observations colored according to the timestamp. The bottom panel displays the same data as a 2D heatmap with time relative to mid-transit on the y-axis. The color scale in this case encodes the flux. The shadow of KELT-18~b if it were aligned is traced by the grey dashed lines. Black rows correspond to gaps in the time series; the narrow hourly bands correspond to etalon calibration images and the large band near +1.8 hrs is when the tip/tilt guiding system failed.}
    \label{fig5:localccfs}
\end{figure*}

We observed a transit of KELT-18~b on UT May 22, 2023 with the Keck Planet Finder \citep[KPF;][]{Gibson2016, Gibson2018, Gibson2020}. KPF is a newly commissioned, optical (445--870~nm), high-resolution ($R \sim 98,000$), fiber-fed, ultra-stabilized radial velocity system on the Keck I telescope at W. M. Keck Observatory. Our observations began 25~min before transit ingress and continued until 2~hours after transit egress, only being interrupted by hourly calibration exposures (described below) and a $\sim$20~min window near transit egress during which issues with the tip/tilt system prevented precise fiber positioning on the stellar PSF.

We chose a fixed 600~sec exposure time to balance averaging over p-mode oscillations (14~min from the scaling relations of \citealt{Brown1991, Kjeldsen1995}) with temporal resolution of the transit, while reaching a spectral signal-to-noise ratio (S/N) of at least 100 (typical values were $130 \pm 11$ in the green channel and $140 \pm 13$ in the red). The KPF ``SKY'' fiber collected background sky contamination from a position offset several arcsec from KELT-18. We simultaneously acquired broadband Fabry-P\'erot etalon spectra in the ``CAL'' fiber to track instrumental drift, and periodically (once per $\sim$hour) took a single internal frame with etalon light in the ``SKY'', ``CAL'', and science fibers as an additional sanity check on drift. We observed a stable linear drift as traced by the simultaneous etalon spectra of $-0.92\pm0.01$~{\ms} per hour in the green channel and $-0.36\pm0.02$~{\ms} per hour in the red channel. This was well-matched by the hourly all-etalon RVs across each fiber. As a result, we drift-corrected our stellar spectra by Doppler-shifting the derived CCFs by estimated drift using our linear fit to the simultaneous etalon RVs.

We independently extracted 1D stellar spectra from each of the three science ``slices'' using the public KPF data reduction pipeline (DRP)\footnote{\href{https://github.com/Keck-DataReductionPipelines/KPF-Pipeline/}{https://github.com/Keck-DataReductionPipelines/KPF-Pipeline/}}. Wavelength calibration was performed for each spectral order using a state-of-the-art laser frequency comb (for $\gtrsim 490$~nm) and a ThAr lamp (for $\lesssim 490$~nm) using calibration frames taken during the standard KPF calibration sequences performed that day. We used the F9 ESPRESSO mask \citep[e.g.][]{Pepe2002} to derive cross-correlation functions \citep[CCF;][]{Baranne1996} for each spectral order. This dataset was obtained before a significant charge transfer inefficiency (CTI) in one of the green CCD amplifiers was diagnosed using solar data \citep{Rubenzahl2023:SoCal}. Because of this, the green CCD readout utilized the original four-amplifier scheme and was thus affected by significant CTI. We masked the flux corresponding to the quadrant of the raw 2D image read by the affected amplifier when deriving CCFs. This affects half of the bluest 20 orders (roughly 445--530~nm). Fortunately, about 78\% of the affected wavelengths also appear in the ``good'' amplifier of the subsequent order, so much of the spectral information is still contained in the final 1D spectrum. The CCFs from each slice were combined in a weighted sum, taking the weights to be proportional to the total flux in each slice from a representative high S/N spectrum. We repeated the same process across all orders, and then again across the green and red CCDs to obtain the final CCF for each observation. We also calculated the unweighted summed CCF to derive photon-noise uncertainties, which we scaled by the relative total flux in the weighted vs.\ unweighted CCF to yield appropriate uncertainties in each CCF.

We independently verified the systemic velocity reported by M17. We measured this by fitting the CCF of each of the three science traces, for each out-of-transit spectrum. The result was $-11.7 \pm 0.1$~{\kms}, in agreement with $-11.6 \pm 0.1$~{\kms} from M17. We found a slightly smaller value of $-11.3 \pm 0.1$~{\kms} by comparing the HIRES spectra used in Section~\ref{sec5:systemprops} to a telluric model \citep{Kolbl2015}. \textit{Gaia} \citep{gaiadr3} reports a smaller but less certain $-10.83 \pm 0.46$~{\kms} for KELT-18. We adopt the $-11.7 \pm 0.1$~{\kms} value from our KPF spectra for our analysis.

\section{Obliquity of KELT-18 b}\label{sec5:obliquity}

\subsection{Reloaded Rossiter-McLaughlin Modeling}
To measure the obliquity of KELT-18~b, we applied the Reloaded Rossiter-McLaughlin technique \citep{Cegla2016} to our KPF spectra. We breifly summarize the process here.

First, we transformed into the stellar rest frame by Doppler shifting the CCFs by our measured systemic velocity and by the expected Keplerian RV induced by KELT-18~b's orbital motion (using the planet's mass $M_p$ from M17). The aligned CCFs were then normalized to a continuum value of 1. We then created a stellar template, CCF$_\text{out}$, by averaging the out-of-transit normalized CCFs. CCF$_\text{out}$ describes the unperturbed average stellar line profile of KELT-18. To isolate the shadow of KELT-18~b, we subtracted each in-transit observation (CCF$_\text{in}$) from the template to obtain the local line profile within the planet's shadow, CCF$_\text{loc}~=~$CCF$_\text{out}~-~$CCF$_\text{in}$. Since we normalized the CCFs, we multiplied each CCF$_\text{loc}$ by the calculated flux at that time according to a white-light synthetic transit light curve model, integrated over the exposure time of each observation. This placed each CCF$_\text{loc}$ on the appropriate flux scale.

The resulting CCF$_\text{loc}$ time series is shown in Figure~\ref{fig5:localccfs}. Each CCF$_\text{loc}$ is fit with a Gaussian profile using \texttt{curve\_fit} from \texttt{scipy} \citep{scipy}, where the continuum, amplitude (i.e., depth), width, and centroid are free parameters. The centroid corresponds to the flux-weighted integrated stellar velocity profile within the shadow of the transiting planet; i.e., the local RV. The local RV is modelled by Eq. 9 in \citep{Cegla2016},
\begin{equation}\label{eq:locrv}
    \text{RV}_\text{loc} = \frac{\int I(x, y)  v_\text{stel}(x, y) dA}{\int I(x, y) dA},
\end{equation}
where $I$ is the limb-darkened intensity, the integral is over the patch of star within the planet shadow, and $v_\text{stel}$ is the stellar velocity field
\begin{equation}\label{eq:vstel}
    v_\text{stel} = x_\perp \veq \sini (1 - \alpha y_\perp'^2),
\end{equation}
where $\alpha$ is the relative differential rotation rate (the difference in rotation rates at the poles compared to the equator, divided by the equatorial rotation rate). For the Sun, $\alpha = 0.27$. We used the same coordinate definitions for $(x_\perp, y_\perp')$ as in \citet{Cegla2016}, but we adapted the implementation of the integral for improved resolution. Instead of defining a Cartesian grid of points $(x_k, y_k)$ centered on the planet's shadow spanning $-R_p/R_\ast$ to $+R_p/R_\ast$ and only keeping the points in the grid which satisfied $x^2 + y^2 < (R_p/R_\ast)^2$, we generated a ``grid'' of $N$ points $(x_k, y_k)$ according to the sunflower pattern,
\begin{equation}\label{eq:sunflower}
\begin{aligned}
    r_k = \sqrt{\frac{k - 1/2}{N-1/2}}, \; \theta_k = k(2\pi\phi), \\
    x_k = r_k\cos\theta_k,\; y_k = r_k\sin\theta_k,
\end{aligned}
\end{equation}
where $\phi = (1+\sqrt{5})/2$ is the golden ratio. The result is a set of points $(x_k,\,y_k)$ uniformly spaced over a circle of radius unity. The points can then be scaled to $R_p/R_\ast$ and centered at the position of the planet to quickly obtain a densely packed grid of points for which each point represents the same projected area $dA$ of the stellar photosphere. The improved resolution of this grid at the shadow and disk limbs helped to reduce artifacts during ingress/egress, and greatly boosted performance when simulating full line profiles for stars with surface inhomogeneities 
\citep{RubenzahlThesis}
.

\begin{figure*}
    \centering
    \includegraphics[width=\textwidth]{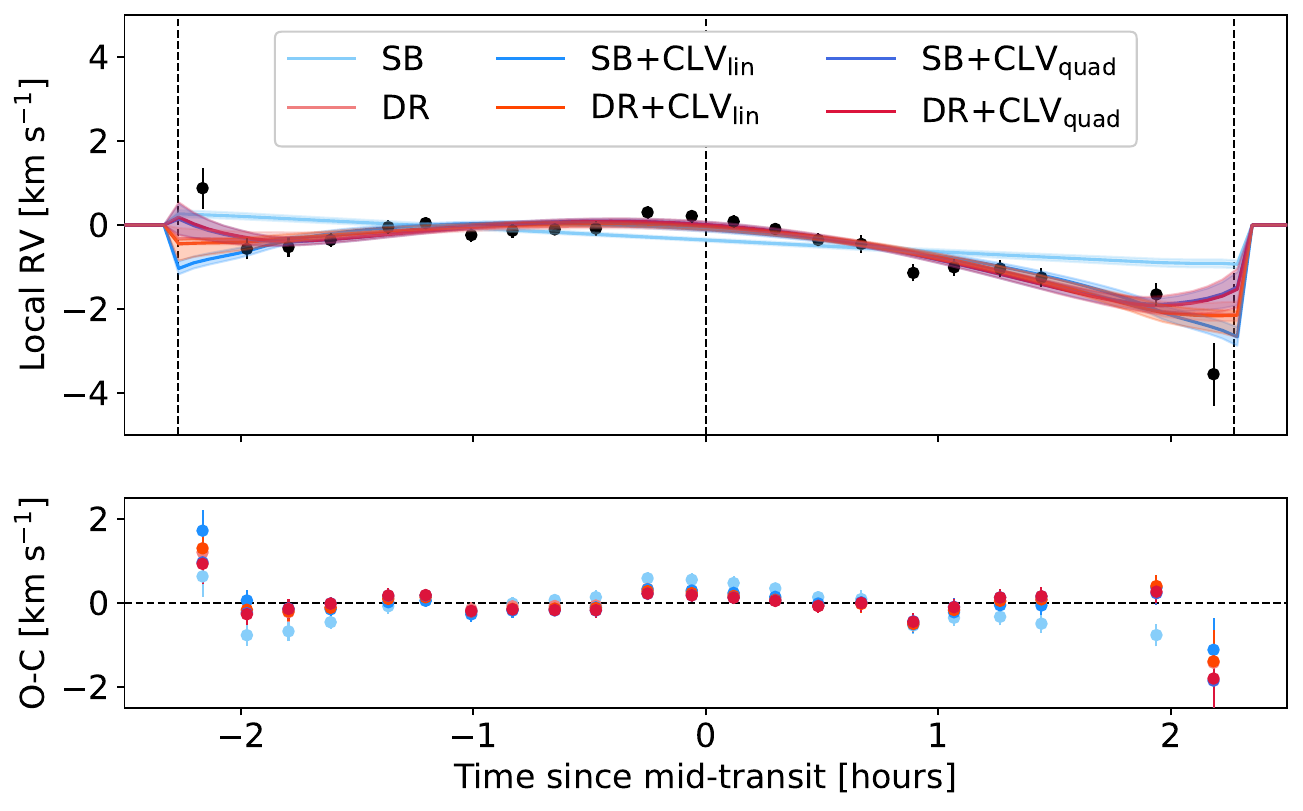}
    \caption{The extracted local RVs and the best-fit RRM models (SB=solid body, DR=differential rotation, CLV=center-to-limb variations as a linear (lin) or quadratic (quad) effect in $\avgmu$). The solid lines are the MAP model while the shaded regions cover the 16th--84th percentile of posterior distribution of models. Thee bottom panel compares the residuals between the data and the MAP fit for each model.}
    \label{fig5:rrmfit}
\end{figure*}

To fit the local RVs, we modified the \texttt{radvel} package \citep{radvel} to accept a new function that computes Eq.~\ref{eq:locrv} for each observation. The \texttt{radvel} framework automatically enabled us to perform maximum a-posteriori (MAP) fitting, MCMC sampling using \texttt{emcee} \citep{emcee}, and model comparison with the BIC and AIC. We tested several different models: solid body (SB) rotation vs. differential rotation (DR), and with/without center-to-limb variations (CLVs); see \citet{Doyle2023} for more details. The former is a matter of fixing $\alpha$ to zero (SB) or letting it float (DR), while the latter requires adding an additional term to Eq.~\ref{eq:locrv} of the form
\begin{equation}\label{eq:clv}
    v_\text{conv} = \sum_{i=0}^n c_i \avgmu^i.
\end{equation}
This polynomial in the intensity-weighted center-to-limb position $\avgmu$ is a good model for the velocity field introduced by granulation, which is azimuthally symmetric around the disk and varies with center-to-limb position as the line-of-sight intersects the tops of granules at disk-center and the sides of granules at the limb \citep{Cegla2016}. Since CCF$_\text{loc}$ has the out-of-transit template subtracted, the net convective blueshift integrated across the full stellar disk has also been removed from the data. Consequently, $c_0$ must be constrained according to Eq. 13 in \citet{Cegla2016}. The additional model parameters are thus $c_1$ for a linear ($n=1$) CLV and $(c_1, c_2)$ for a quadratic ($n=2$) CLV.

\subsection{MCMC Sampling}

We tested a suite of models within the RRM framework corresponding to each combination of SB or DR, and no CLV, linear CLV (CLV$_\text{lin}$), and quadratic CLV (CLV$_\text{quad}$). The free parameters in all models were the sky-projected obliquity $\lambda$, the projected rotational velocity $\vsini$, the sine of the stellar inclination $\sini$, and the impact parameter $b$. Models with DR include the degree of differential rotation $\alpha$, and models with CLVs include either $c_1$ (linear) or $c_1$ and $c_2$ (quadratic). We allowed for anti-solar differential rotation by permitting $\alpha$ to be negative, with a uniform prior over $(-1, 1)$.

To improve the sampling efficiency, we make the change of coordinates for our fitting basis into a polar coordinate system with $\lambda$ as the azimuthal angle and $\sqrt{\vsini}$ as the radial dimension. The parameters for the fit are thus $\sqrt{\vsini}\cos\lambda$, $\sqrt{\vsini}\sin\lambda$, $\sin\istar$, $b$, $\alpha$, and the CLV coefficients.

Because of the suspected polar orientation of the transit chord, we placed an informed Gaussian prior on $\vsini$ of $10.4 \pm 1$~{\kms} based on our analysis of the spectroscopic $\vsini$ (Section~\ref{sec5:systemprops}). We also found that this prior, in conjunction with a prior on $\iorb$ based on previous transit fits (Table~\ref{tab5:systemprops}), was necessary to discourage the sampler from wandering to solutions of extremely low $\vsini$ ($\sim$1~{\kms}) with (unrealistic) grazing transits. The final distributions for $\lambda$ were unaffected by the exact boundaries chosen for these priors. We note that \citet{Maciejewski2020} found values of $\iorb$ ($82.90^{+0.62}_{-0.54} {}^\circ$) and $a/R_\ast$ ($4.36^{+0.11}_{-0.09}$) that were significantly discrepant with those measured by M17. We tried setting a prior on $\iorb$ to this value and found the MCMC to both not converge and produce a bimodal $\vsini$ posterior around 2~{\kms}, which is highly inconsistent with the observed width of lines in the KPF spectra and the HIRES, TRES, and APF spectra of M17.

We first found the MAP solution for each model using \texttt{scipy.optimize.minimize} \citep{scipy}. This best-fit solution was used as the initial location (plus a small Gaussian perturbation) for a MCMC exploration of the posterior. We ran \texttt{emcee} \citep{emcee} as implemented in \texttt{radvel} \citep{radvel} with 8 ensembles of 32 walkers each for a maximum of 10,000 steps, or until the Gelman--Rubin statistic~\citep[G--R;][]{Gelman2003} was $< 1.001$ across the ensembles~\citep{Ford2006}. In all cases, the G-R condition for convergence was satisfied and the sampler was terminated with a typical total number of posterior samples around 100,000. A second and final MAP fit was then performed using the median parameter values from the MCMC samples. The MAP fit for each model is plotted over the RV$_\text{loc}$ time series in Figure~\ref{fig5:rrmfit}, and Table~\ref{tab5:rrmparams} lists our derived best-estimates for each parameter.

\subsection{Center-to-Limb Variations}

\begin{figure*}
    \centering
    \includegraphics[width=0.9\textwidth]{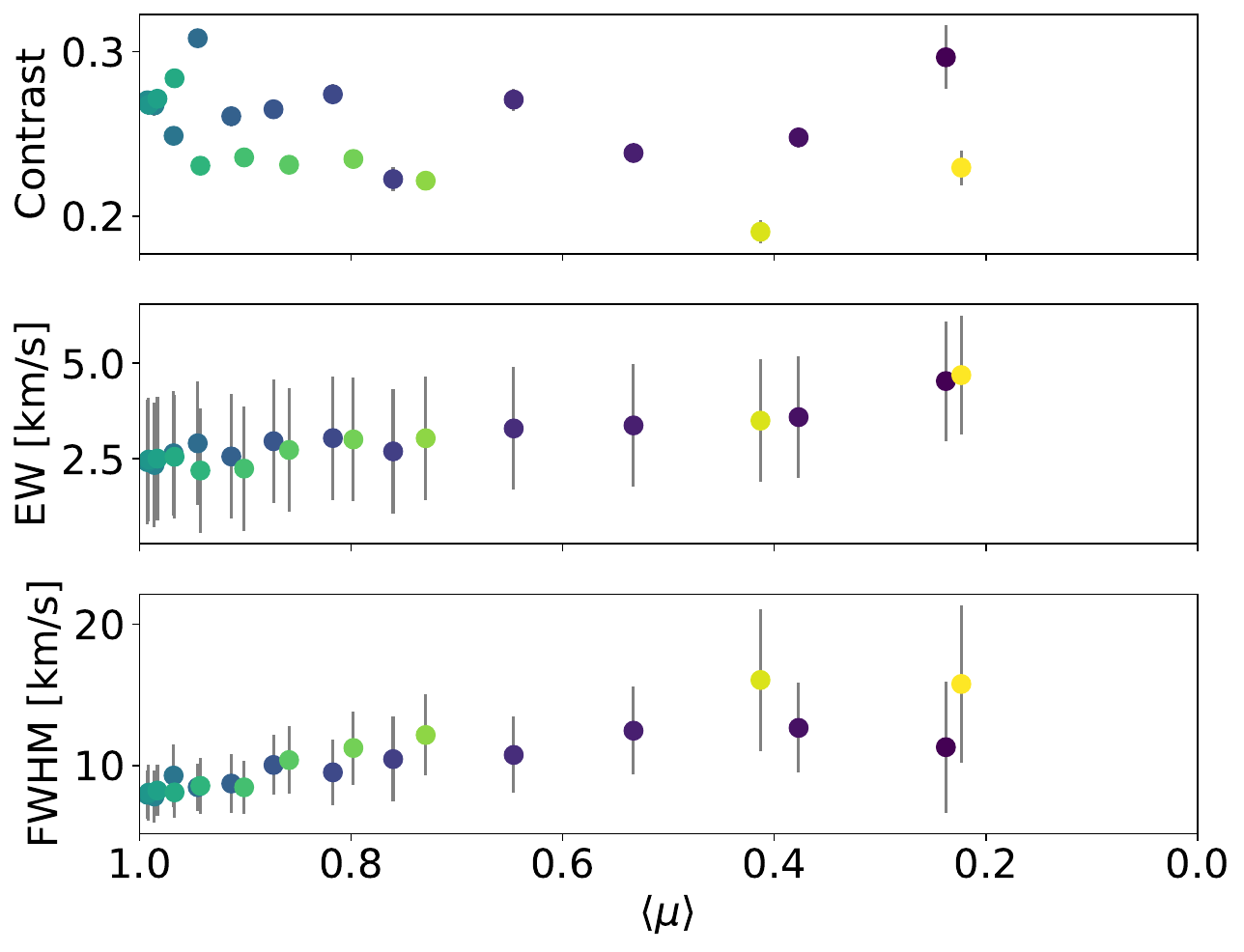}
    \caption{The contrast (top), equivalent width (middle), and FWHM (bottom) of CCF$_\text{loc}$ as a function of the flux-weighted center-to-limb position $\avgmu$. The color scale is the same used in Figure~\ref{fig5:localccfs} based on the observation timestamps (purple=ingress, green=mid transit, yellow=egress).}
    \label{fig5:clv}
\end{figure*}

Fig.~\ref{fig5:clv} plots the depth, equivalent width (EW), and full-width-at-half-maximum (FWHM) of the CCF$_\text{loc}$ as a function of $\avgmu$. A strong trend in the FWHM (Pearson correlation coefficient R=-0.87, p-value of $1.5\times10^{-7}$) and the EW (R=-0.94, p-value $1.6\times10^{-10}$) can be seen from center to limb, with the local line profiles narrowest at disk center and widening towards disk limb. The RV$_\text{loc}$ time series (Figure~\ref{fig5:rrmfit}) likewise has significant curvature with a maximum (minimum blueshift) at disk center (mid-transit) and minimum (maximum blueshift) at disk limb (ingress/egress). The CCF$_\text{loc}$ contrast is only marginally correlated with CLV position during the first half of the transit (R=0.66, p-value 0.04) and is uncorrelated during the second half of the transit (R=0.03, p-value 0.94).

\begin{figure*}
    \centering
    \includegraphics[width=\textwidth]{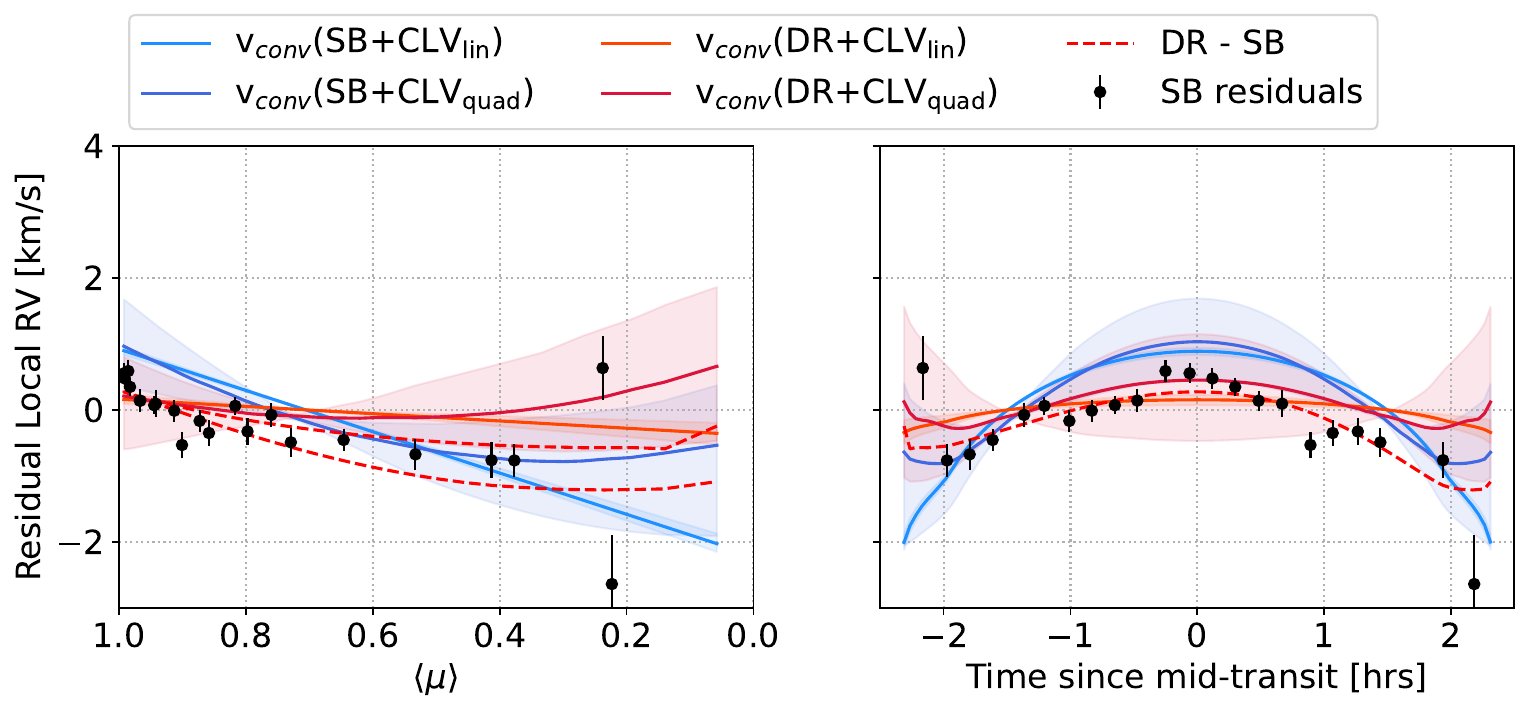}
    \caption{The best-fit convective velocity profiles within the patches of star occulted by KELT-18~b as a function of center-to-limb position (left) and time (right). The $v_\text{conv}$ for models with DR are consistent with zero, i.e. the DR models are primarily fit by extreme DR. The velocity contribution from DR alone is shown by the dashed red line, which plots the difference between the DR-only and SB-only models. Residuals (data minus model) to the SB-only model are plotted in black in both panels; these points represent the amount of local RV that must be coming from either DR or CLVs and are shown as a representation of what the DR and/or CLV components of the other models are effectively fitting}.
    \label{fig5:clvmodel}
\end{figure*}

Only CLVs can simultaneously explain the local RV curvature and the change in line width. While differential rotation alone can reproduce the observed curvature in the local RV time series, it cannot explain the large (factor of two) increase in FWHM from disk center to limb. \citet{Beeck2013b} showed using spectral line synthesis in 3D radiative magnetohydrodynamical (MHD) simulations that surface-layer granulation causes the width of spectral lines to increase towards disk limb, an effect that is relatively muted for GKM stars but is significant for F-type stars. The F3 V star in their simulation showed increases to the FWHM of iron lines by a factor of 1.5--2 from disk center to $\mu = 0.2$. This effect is caused by the horizontal flows, which dominate the line-of-sight velocity at disk limb, having roughly three times as high a velocity dispersion compared to the vertical velocity \citep{Beeck2013a}. They also found line cores to have a net convective blueshift about 500~{\ms} less (i.e., larger RV) at disk limb than at disk center. We see the opposite in the modelled CLV vs. $\mu$ (Fig.~\ref{fig5:clvmodel}) for the SB scenario, whereas in the DR scenario we find negligible convective velocities. 


\input{rrmparams}
\begin{figure*}
    \centering
    \includegraphics[width=\textwidth]{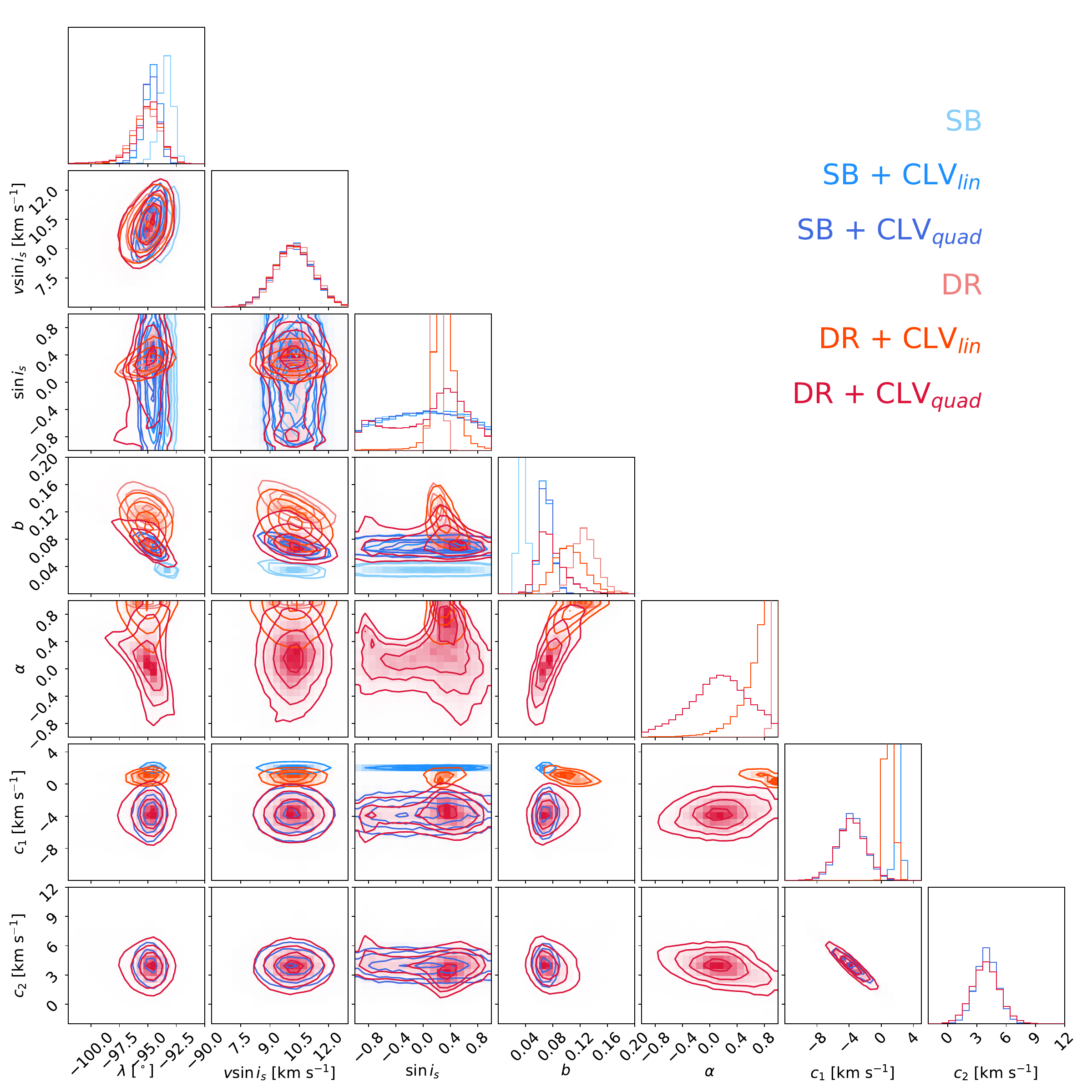}
    \caption{The posterior distributions for all RRM parameters. The models with DR are colored red while SB models are colored blue, and models with no, linear, or quadratic CLVs are given progressively darker colors. All models generally agree on the value of $\lambda$, while models with DR tend towards extreme values of $\alpha$ at near-polar $\istar$. This degeneracy arises because of the tight prior on $\vsini$ (2nd column), which all models are confined to.}
    \label{fig5:cornerplot}
\end{figure*}

\subsection{Model Comparison} \label{sec5:modelcomparison}

We computed the Bayesian information criterion (BIC) and Akaike information criterion (AIC) for each model. Their relative values to the minimum are listed in Table~\ref{tab5:rrmparams}. The only model that is confidently ruled out is SB rotation with no CLV, at $\Delta$BIC=74 and $\Delta$AIC=73. The slightly preferred model is SB rotation with CLVs quadratic in $\avgmu$. The models with DR all have $\Delta$BIC and $\Delta$AIC $<5$, and thus are statistically similar descriptions of the model.

The model with DR alone requires an extremely high values of $\alpha$, in the 0.9--1 range. While it is not impossible for the stellar poles to rotate at just 10\% the rate of the equator, it seems extremely unlikely for a star to have a rotational shear of this magnitude. Slowly rotating ($\vsini \lesssim 50$~{\kms}) F-type stars do commonly show signs of differential rotation ($\alpha \geq 0.1$), whereas rapid rotators do not \citep{Reiners2003}. Our analysis of the TESS photometry in Section~\ref{sec5:rotation-period} more likely places KELT-18 in the former category of slow rotators. Interestingly, the DR model with linear CLVs finds a smaller $\alpha=0.8_{-0.27}^{+0.14}$, and the flexibility of the quadratic CLV model results in an unconstrained $\alpha$, as the two effects are degenerate. In contrast, the SB models rely on strong CLVs to generate the measured curvature (Fig.~\ref{fig5:clvmodel}). It is therefore ambiguous from the RV$_\text{loc}$ time series alone whether the curvature is coming from a significant differential rotation, CLVs, or a mixture of the two. The difference between these two cases is greatest at low $\langle\mu\rangle$, i.e. near the disk limb. This is where CLV effects are strongest whereas DR is only affected by the subplanet stellar latitude. However, the data near disk limb (i.e. near ingress/egress) are the lowest S/N observations, weakening their utility as a discriminatory lever-arm.

If the star is differentially rotating, the varying line-of-sight rotational velocities as a function of stellar latitude break the $\sini$ degeneracy, allowing an independent constraint on the stellar inclination. In the DR-only model, the resulting stellar inclination is $\istar = 11.8_{-3.2}^{+3.3}$. This value is in agreement with the 5--30$^\circ$ range expected from the estimated rotation period and known $R_\ast$ (Section~\ref{sec5:rotation-period}). Consequently, the true 3D obliquity can be determined via the equation
\begin{equation}\label{eq:trueobliquity}
    \cos\psi = \cos\istar \cos\iorb + \sini \sin\iorb \cos\lambda.
\end{equation}
The corresponding values of $\psi$ for each of fitted models are listed in Table~\ref{tab5:rrmparams}. For the SB models, the posterior in $\sini$ is unconstrained (uniform in 0 to 1) so this represents the maximally uncertain value of $\psi$. For instance, if we adopt $\lambda = \bestlam$ from the best-fitting SB+CLV$_\text{quad}$ model, and take $\istar$ to be isotropic (uniform in $\cos\istar$), then $\psi = 93.8_{-1.8}^{+1.6}{}^\circ$. If instead we take $\istar$ to be isotropic within 5--30$^\circ$ as suggested from photometric analyses, we get $\psi = \bestpsi$. Our estimated obliquity measurements are consistent within $1\sigma$ across all models, thus we adopt the SB+CLV$_\text{quad}$ model as our preferred fit as it is both the best-fitting model by the $\Delta$BIC and $\Delta$AIC tests, is more physically justified than the $\alpha \sim 0.9$ models, and utilizes photometric constraints on the stellar inclination rather than the values from the RRM posteriors.

\section{Orbital Dynamics}\label{sec5:dynamics}

KELT-18~b joins the population of close-in planets orbiting hot stars in polar orbits \citep{AlbrechtReview}. There are no other known transiting planets in the system \citep{Maciejewski2020}, though there is a stellar companion, KELT-18~B (Section~\ref{sec5:companion}). It is likely that KELT-18~B influenced the orbital evolution of KELT-18~b. In this section we examine possible mechanisms.


Hierarchical triple systems such as KELT-18, KELT-18~b, and KELT-18~B will experience von-Zeipel-Kozai-Lidov (ZKL) oscillations if the mutual inclination between the two orbiting bodies is larger than $39.2^\circ$ \citep{Naoz2016}. ZKL oscillations are usually suppressed for planets on HJ-like orbits because of the general relativistic precession of the argument of periapsis. However ZKL oscillations may have taken place early in the system's history if KELT-18~b formed at a further orbital distance. \citet{Fabrycky2007} showed that such a scenario can lead to HEM, producing the HJ we see today in a close-in, highly misaligned orbit.

\begin{figure*}
    \centering
    \includegraphics[width=0.95\textwidth]{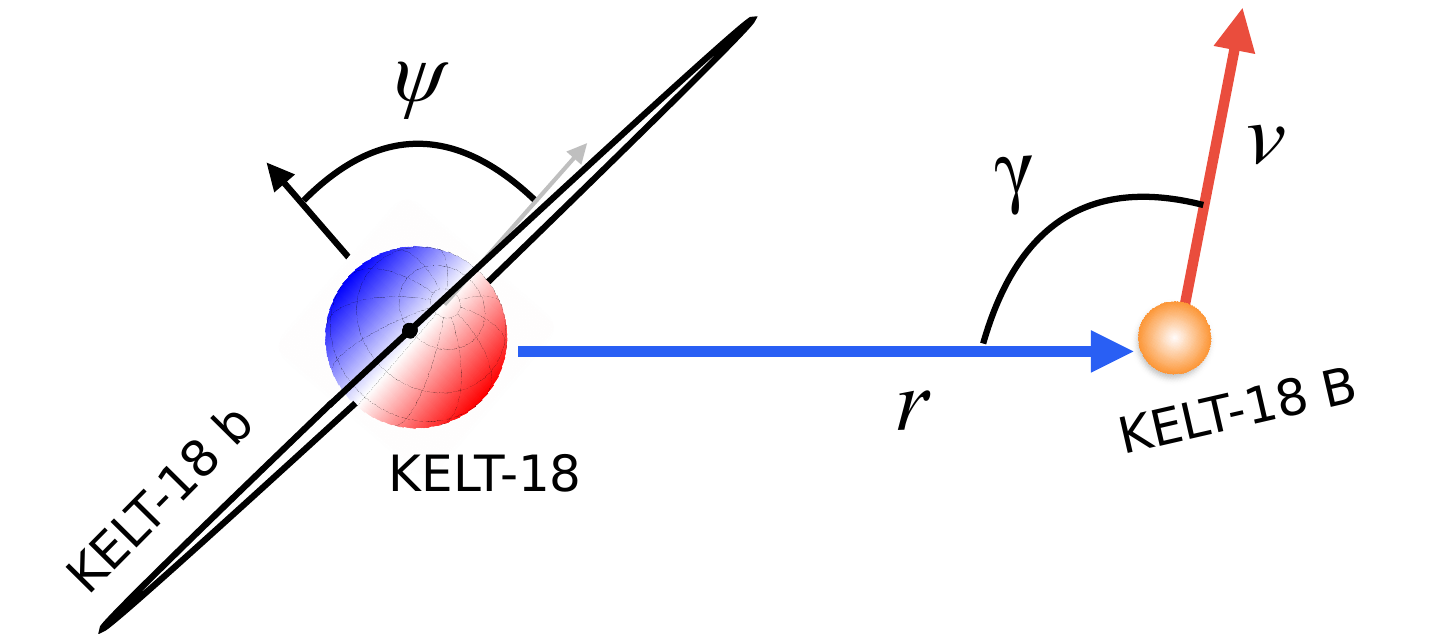}
    \caption{On-sky geometry of the KELT-18 system. KELT-18 and KELT-18~b (and its orbit) are drawn to scale in size and relative orientation. The black arrow denotes the normal to KELT-18~b's orbital plane and the grey arrow denotes the rotation axis of KELT-18; the angle between these in 3D space is the stellar obliquity, $\psi$, while $\lambda$ is the sky-projection of this angle (i.e., the angle on this page) and is independent of the stellar inclination. The separation between KELT-18 and KELT-18~B is not to scale and is drawn at an arbitrary orientation on-sky. KELT-18 is colored according to its rotational velocity profile (with no differential rotation) and lines of latitude/longitude are drawn to illustrate the orientation of the pole ($\istar = 30^\circ$). The angle between the binary star position vector ($\vec{r}$, blue arrow) and relative proper motion ($\vec{v}$, red arrow), $\gamma$, is labelled. }
    \label{fig5:onsky}
\end{figure*}

The \textit{Gaia} astrometry enables a constraint on the inclination of KELT-18~B's orbit. The angle between the vector connecting the astrometric positions of the two stars ($\vec{r}$) and the difference in velocity vectors ($\vec{v}$), called $\gamma$ \citep[e.g.,][see Fig.~\ref{fig5:onsky}]{Tokovinin2015, Hwang2022}, encodes information about the companion's orbital inclination (though is degenerate with eccentricity). If $\gamma \sim 0^\circ$ or $180^\circ$, then the companion's orbit is viewed edge-on. Otherwise, the companion's orbit may be viewed face-on or at some intermediate inclination. Since KELT-18~b transits, we know it has $\iorb \sim 90^\circ$, so $\gamma$ can test for mutual (mis)alignment. We calculated $\gamma = 95.2 \pm 6.4^\circ$, which is most consistent with a low orbital inclination for a circular KELT-18~B and thus a large mutual inclination (Fig.~\ref{fig5:onsky}).

For ZKL to excite KELT-18~b's orbit, it must have formed far away from its host star. By equating the timescales for GR precession and ZKL oscillations, one can solve for the minimum orbital separation at which ZKL oscillations are not quenched by GR \citep[Eq.~4 of][]{Dong2014}. We computed this value for the KELT-18 system and obtained $\gtrsim 6.1\pm0.26$~AU. In other words, if KELT-18~b was born beyond $6.16\pm0.26$~AU, for instance via traditional core accretion beyond the ice-line, then it could have plausibly migrated to its current orbit via ZKL-induced HEM. B22 calculated typical minimum formation distances of 0.5--10~AU across the broader population of HJs for binary star induced ZKL HEM migration, which conspicuously aligns with the peak in cold Jupiter occurrence around 1--10~AU \citep{Fulton2021}. Future studies of KELT-18~b's atmosphere via transmission spectroscopy (Householder, A., Dai, F., et al. in prep) will seek additional evidence for KELT-18~b's birth conditions by measuring its inventory of refractory and volatile elements, the fingerprints of the original planetary building blocks \citep{Lothringer2021}.

It may also be the case that KELT-18~b formed in a protoplanetary that was primordially misaligned, as can be the case when an outer stellar companion is involved \citep{Batygin2012}. Alternatively, the stellar companion may have torqued the outer regions of the protoplanetary disk into a misalignment, producing a broken protoplanetary disk which itself can play the role of an outer perturber in exciting large stellar obliquities \citep{EpsteinMartin2022}. \citet{Vick2023} showed that such an initial configuration, in which the star has a disk-induced nonzero obliquity relative to the proto-HJ before ZKL oscillations induced by the stellar companion initiate HEM, the final obliquity distribution of the HJ is broadly retrograde with a peak near polar orbits. This is in contrast with the classical picture of ZKL starting with initially aligned planetary orbits, which produce HJs with a bimodal obliquity distribution near 40$^\circ$ and 140$^\circ$ \citep{Anderson2016}. Thus it may be that the orbit of KELT-18~b was already misaligned with the star's rotation before undergoing HEM into its present day HJ orbit, the result of which is an orbit with $\psi \sim 90^\circ$ rather than 40$^\circ$ or 140$^\circ$.

\section{Conclusion}
\label{sec5:conclusion}

We have presented the first science results from KPF on the Keck-I telescope: a transit of the inflated ultra-hot Jupiter KELT-18~b. We found the orbit to be nearly perpendicular to the stellar equatorial plane: $\psi = \bestpsi$. This result is robust to model choice and is largely constrained by the tight posterior on $\lambda = \bestlam$ and the relatively low value of $\vsini$. Taken in context with the binary stellar companion, which we find to be on a likely bound orbit that could be orthogonal to KELT-18~b's orbit, a history of ZKL-induced HEM is plausible if KELT-18~b formed beyond about 6~AU from its host star. Our main observational takeaways are as follows:
\begin{itemize}
    \item We searched the available TESS photometry for clues as to the rotation period of the host star and found evidence for modulation around $\sim$5~days. We did not see variability at 0.707~days as previously reported by M17 using KELT photometry. Both values are consistent with the tendency for F-type stars to have $< 8$~day rotation periods, and when combined with the measured $\vsini$ imply a near pole-on viewing geometry ($\istar \lesssim 30^\circ$).

    \item The stellar neighbor KELT-18~B is highly likely to be a bound companion, based on \textit{Gaia} DR3 astrometry. Its orbit is also likely orthogonal to KELT-18~b's orbit, based on the angle between the on-sky position vector and proper motion vectors.

    \item We observed evidence of CLVs, as traced by the FWHM of the local line profile beneath the planet's shadow. The FWHM increased towards the disk limb by nearly a factor of two, in agreement with previous 3D MHD simulations of velocity flows in the near-surface layers of F-type stellar atmospheres. 
    
    \item We modelled the centroid of the local line profile using the RRM technique and found that either strong differential rotation ($\alpha=0.9$) or CLVs are needed to explain the curvature in the local RV time series. However, all of the models produce consistent, well-constrained posteriors for the sky-projected obliquity. Ambiguity between DR and CLVs is a common challenge of the RRM technique \citep[see e.g.][]{RoguetKern2022, Doyle2023} and is complicated by the uncertainty in the stellar inclination (and thus the stellar latitudes occulted). A firm detection of a stellar rotation period from additional photometry would enable a better constraint of the degree of DR needed to explain the data. As it stands, the polar transiting geometry requires a low (near pole-on) stellar inclination in order to generate the observed curvature in the local RV timeseries, with maximum blueshift occulted at ingress/egress and near-zero velocity occulted at mid-transit. An edge on stellar inclination would produce the opposite effect, since the lowest velocity latitudes at the poles would instead be occulted at ingress/egress, while the maximum velocity latitude (the equator) would be occulted at mid-transit. If KELT-18 does have an edge-on stellar inclination, then DR would be inconsistent with the data and CLVs would be strongly favored.

    \item The 3D orbital geometry of the KELT-18 system is explainable by a history of ZKL-induced migration, providing support for the HEM formation pathway for HJs. Future work will further test this by inventorying the elemental abundances in KELT-18's atmosphere, connecting the planet to the disk in which it formed.

\end{itemize}

\section{Acknowledgements}

We are grateful to Heather Cegla and Michael Palumbo for illuminating discussions on center-to-limb variations. Some of the data presented herein were obtained at Keck Observatory, which is a private 501(c)3 non-profit organization operated as a scientific partnership among the California Institute of Technology, the University of California, and the National Aeronautics and Space Administration. The Observatory was made possible by the generous financial support of the W. M. Keck Foundation. Keck Observatory occupies the summit of Maunakea, a place of significant ecological, cultural, and spiritual importance within the indigenous Hawaiian community. We understand and embrace our accountability to Maunakea and the indigenous Hawaiian community, and commit to our role in long-term mutual stewardship. We are most fortunate to have the opportunity to conduct observations from Maunakea. 

R.A.R. acknowledges support from the National Science Foundation through the Graduate Research Fellowship Program (DGE 1745301). A.W.H.\ acknowledges funding support from NASA award 80NSSC24K0161 and the JPL President's and Director's Research and Develop Fund. This paper made use of data collected by the TESS mission and are publicly available from the Mikulski Archive for Space Telescopes (MAST) operated by the Space Telescope Science Institute (STScI). All the {\it TESS} data used in this paper can be found in MAST: \dataset[10.17909/t9-nmc8-f686]{http://dx.doi.org/t9-nmc8-f686} \citep{mastdoi}. This research was carried out, in part, at the Jet Propulsion Laboratory and the California Institute of Technology under a contract with the National Aeronautics and Space Administration and funded through the President’s and Director’s Research \& Development Fund Program.

\facility{Keck:I (KPF) \citep{Gibson2020}, TESS \citep{TESS}} 

\software{
\texttt{astropy}    \citep{astropy1, astropy2, astropy3},
\texttt{corner}     \citep{corner},
\texttt{emcee}      \citep{emcee},
\texttt{lightkurve} \citep{lightkurve},
\texttt{matplotlib} \citep{matplotlib},
\texttt{numpy}      \citep{numpy},
\texttt{pandas}     \citep{pandas},
\texttt{radvel}     \citep{radvel},
\texttt{scipy}      \citep{scipy},
\texttt{SpecMatch-Emp} \citep{smemp}, 
\texttt{SpecMatch-Synth} \citep{Petigura2015},
}

\bibliography{references}
\bibliographystyle{aasjournal}


\appendix

\section{Sector-by-sector rotation period}\label{sec:rot-appendix}

\begin{figure*}[b!]
    \centering
    \includegraphics[width=0.6\textwidth]{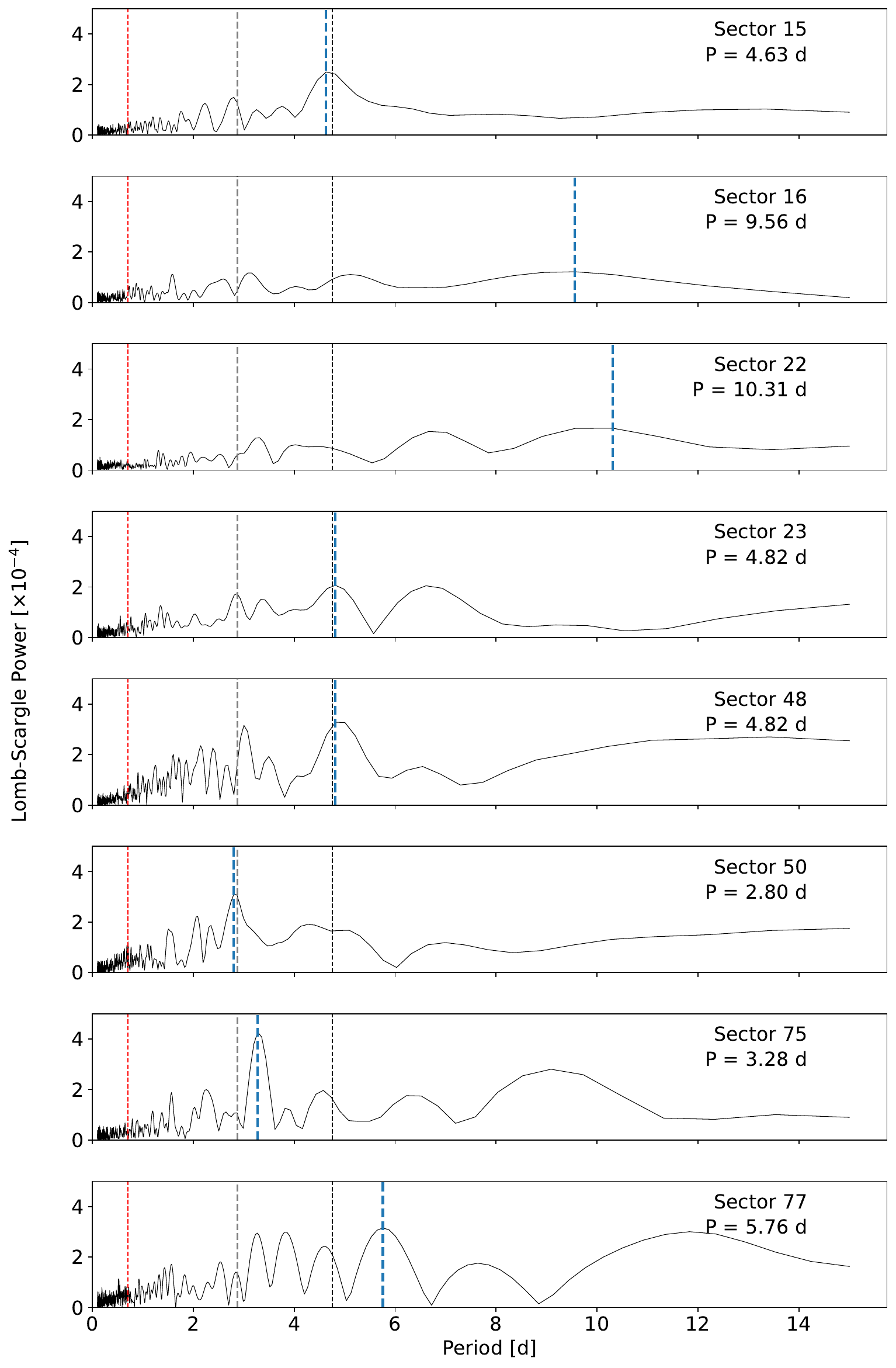}
    \caption{Periodograms, following the same methodology used to produce the periodogram in Figure~\ref{fig5:lightcurve}, for each individual TESS sector. The primary peak near 5 days in the overall periodogram only appears in the individual sectors 15, 23, and 48. Other sectors show different peak structures. As in Figure~\ref{fig5:lightcurve}, the 0.707~d period from M17 is denoted by the red dashed line, KELT-18's orbital period is denoted by the grey dashed line, and the maximum peak is marked by a blue dashed line. The black dotted line gives the maximum peak in the overall periodogram (i.e., the blue line in Figure~\ref{fig5:lightcurve}).}
    \label{fig5:sectorperiodogram}
\end{figure*}

Here we present, in Figure~\ref{fig5:sectorperiodogram}, the sector-by-sector periodogram of the TESS photometry.

\end{document}

%% file: authors.tex
\newcommand{\caltechastro}{Department of Astronomy, California Institute of Technology, Pasadena, CA 91125, USA}
\newcommand{\SSL}{Space Sciences Laboratory, University of California, Berkeley, CA 94720, USA}
\newcommand{\wmko}{W. M. Keck Observatory, Waimea, HI 96743, USA}
\newcommand{\UCO}{UC Observatories, University of California, Santa Cruz, CA 95064, USA}
\newcommand{\ipac}{NASA Exoplanet Science Institute/Caltech-IPAC, MC 314-6, 1200 E. California Blvd., Pasadena, CA 91125, USA}

\author[0000-0003-3856-3143]{Ryan A. Rubenzahl}
\altaffiliation{NSF Graduate Research Fellow}
\affiliation{\caltechastro}

\author[0000-0002-8958-0683]{Fei Dai}
\affiliation{Division of Geological and Planetary Sciences,
1200 E California Blvd, Pasadena, CA, 91125, USA}
\affiliation{Institute for Astronomy, University of Hawai{\okina}i, 2680 Woodlawn Drive, Honolulu, HI 96822, USA}

\author[0000-0003-1312-9391]{Samuel Halverson}
\affiliation{Jet Propulsion Lab, Pasadena, CA 91125, USA}

\author[0000-0001-8638-0320]{Andrew W. Howard}
\affiliation{\caltechastro}

\author[0000-0002-5812-3236]{Aaron Householder}
\affiliation{Department of Earth, Atmospheric and Planetary Sciences, Massachusetts Institute of Technology, Cambridge, MA 02139, USA}
\affil{Kavli Institute for Astrophysics and Space Research, Massachusetts Institute of Technology, Cambridge, MA 02139, USA}

\author[0000-0003-3504-5316]{Benjamin Fulton}
\affiliation{\ipac}

\author[0000-0003-0012-9093]{Aida Behmard}
\altaffiliation{Kalbfleisch Fellow}
\affiliation{American Museum of Natural History, 200 Central Park West, Manhattan, NY 10024, USA}

\author[0009-0004-4454-6053]{Steven R. Gibson}
\affil{Caltech Optical Observatories, Pasadena, CA, 91125, USA}

\author[0000-0001-8127-5775]{Arpita Roy}
\affiliation{Astrophysics \& Space Institute, Schmidt Sciences, New York, NY 10011, USA}

\author[0000-0003-3133-6837]{Abby P. Shaum}
\affiliation{\caltechastro}

\author[0000-0002-0531-1073]{Howard Isaacson}
\affiliation{501 Campbell Hall, University of California at Berkeley, Berkeley, CA 94720, USA}


\author[0009-0008-9808-0411]{Max Brodheim}
\affiliation{\wmko}

\author[0009-0000-3624-1330]{William Deich}
\affil{\UCO}

\author[0000-0002-7648-9119]{Grant M. Hill}
\affiliation{\wmko}

\author[0000-0002-6153-3076]{Bradford Holden}
\affil{\UCO}

\author[0000-0003-2451-5482]{Russ R. Laher}
\affiliation{\ipac}

\author[0009-0004-0592-1850]{Kyle Lanclos}
\affiliation{\wmko}

\author[0009-0008-4293-0341]{Joel N. Payne}
\affiliation{\wmko}

\author[0000-0003-0967-2893]{Erik A. Petigura}
\affiliation{Department of Physics \& Astronomy, University of California Los Angeles, Los Angeles, CA 90095, USA}

\author[0000-0002-4046-987X]{Christian Schwab}
\affiliation{School of Mathematical and Physical Sciences, Macquarie University, Balaclava Road, North Ryde, NSW 2109, Australia}

\author{Chris Smith}
\affiliation{\SSL}

\author[0000-0001-7409-5688]{Gu{\dh}mundur Stef{\'a}nsson}
\affil{Anton Pannekoek Institute for Astronomy, University of Amsterdam, Science Park 904, 1098 XH Amsterdam, The Netherlands}

\author[0000-0002-6092-8295]{Josh Walawender}
\affiliation{\wmko}

\author[0000-0002-6937-9034]{Sharon X.~Wang}
\affiliation{Department of Astronomy, Tsinghua University, Beijing 100084, People's Republic of China}

\author[0000-0002-3725-3058]{Lauren M. Weiss}
\affiliation{Department of Physics and Astronomy, University of Notre Dame, Notre Dame, IN 46556, USA}

\author[0000-0002-4265-047X]{Joshua N.\ Winn}
\affiliation{Department of Astrophysical Sciences, Princeton University, Princeton, NJ 08544, USA}

\author[0000-0002-4265-047X]{Edward Wishnow}
\affiliation{\SSL}

%% file: kelt18systemprops.tex
\begin{deluxetable}{lrrr}
\centering
\tablecaption{Parameters of the KELT-18 System
\label{tab5:systemprops}}
\tablehead{
  \colhead{Parameter} & 
  \colhead{Value} & 
  \colhead{Unit} &
  \colhead{Source} 
}
\startdata 
\textbf{KELT-18} \\
$\Teff$          & $6670\pm 120$             & K         & M17 \\
$M_\ast$         & $1.524^{+0.069}_{-0.068}$ & M$_\odot$ & M17 \\
$R_\ast$         & $1.908^{+0.042}_{-0.035}$ & R$_\odot$ & M17 \\
$u_1$            & $0.337_{-0.010}^{+0.011}$ &           & M17 \\
$u_2$            & $0.3229_{-0.0059}^{+0.0066}$&         & M17 \\
RV$_\text{sys}$  & $-11.7 \pm 0.1$           & {\kms}    & This work \\   
\hline
\textbf{KELT-18 b} \\
$\Porb$         & $2.87169867 \pm 0.00000085$ & days    & I22 \\
$t_c$           & $2458714.17773 \pm 0.00011$ & JD      & I22 \\
$b$             & $0.10^{+0.10}_{-0.07}$      &         & M17 \\ 
$\iorb$         & $88.86^{+0.79}_{-1.20}$     & degrees & M17 \\
$R_p/R_\star$   & $0.08462 \pm 0.00091$       &         & M17 \\
$a / R_\star$   & $5.138^{+0.038}_{-0.078}$   &         & M17 \\
$e$             & $0$                         &         & M17 \\
$M_p$           & $1.18 \pm 0.11$             & M$_J$   & M17 \\
\hline
\textbf{KELT-18 B} \\
$\Teff$          & $3900$                    & K         & M17 \\
$M_\ast$         & $0.575_{-0.026}^{+0.025}$ & M$_\odot$ & B22 \\
sep              & 1082                      & AU        & B22 \\
$\Delta\mu_\text{RA}$ & $0.21\pm0.05$    & mas~yr$^{-1}$ & B22 \\
$\Delta\mu_\text{Decl.}$ & $0.41\pm0.05$ & mas~yr$^{-1}$ & B22 \\
$\Delta$parallax & $0.07\pm0.04$         & mas           & B22 \\
$\Delta$G        & $-5.4$                & mag           & B22 \\
$\ln(\mathcal{L}_1/\mathcal{L}_2)$ & 4.83& ---           & B22 \\
\hline
\enddata
\tablenotetext{}{(M17) \citet{McLeod2017}; (I22) \citet{Ivshina2022}; (B22) \citet{Behmard2022}.}
\end{deluxetable}

%% file: rrmparams.tex
\startlongtable
\begin{deluxetable*}{llllllll}
\tablecaption{Best-fit Parameters and Model Comparison} 
\label{tab5:rrmparams}
\tablehead{
\colhead{Parameter}  & \colhead{Prior} & 
\colhead{SB} & \colhead{SB+CLV$_\text{lin}$} & \colhead{\textbf{SB+CLV$_\text{quad}$}} &  \colhead{DR} & \colhead{DR+CLV$_\text{lin}$} & \colhead{DR+CLV$_\text{quad}$}
}
\startdata
\textbf{Model Parameters} \\
$\lambda$ ($^\circ$) & --- & $-93.4_{-0.6}^{+0.6}$ & $-94.6_{-0.7}^{+0.6}$ & $\bm{-94.8_{-0.7}^{+0.7}}$ & $-95.2_{-1.2}^{+1.2}$ & $-95.0_{-1.2}^{+1.0}$ & $-94.9_{-1.3}^{+1.0}$ \\
$\vsini$ ({\kms}) & Gaussian(10.4, 1) & $10.2_{-1.0}^{+1.0}$ & $10.2_{-1.0}^{+1.0}$ & $\bm{10.2_{-1.0}^{+1.0}}$ & $10.3_{-1.0}^{+1.0}$ & $10.2_{-1.0}^{+1.0}$ & $10.2_{-1.0}^{+1.0}$ \\
$\sini$ & Uniform(0, 1) & $0.03_{-0.63}^{+0.6}$ & $0.06_{-0.64}^{+0.59}$ & $\bm{0.0_{-0.61}^{+0.6}}$ & $0.2_{-0.05}^{+0.06}$ & $0.29_{-0.12}^{+0.15}$ & $0.22_{-0.81}^{+0.38}$ \\
$b$ & Uniform(-1, 1) & $0.035_{-0.005}^{+0.006}$ & $0.069_{-0.008}^{+0.01}$ & $\bm{0.07_{-0.009}^{+0.009}}$ & $0.127_{-0.016}^{+0.018}$ & $0.104_{-0.021}^{+0.023}$ & $0.075_{-0.015}^{+0.023}$ \\
$\alpha$ & Uniform(-1, 1) &  --- & --- & \textbf{---} &$0.98_{-0.03}^{+0.01}$ & $0.8_{-0.27}^{+0.14}$ & $0.15_{-0.45}^{+0.42}$ \\
$c_0$ ({\kms}) & Uniform(-10, 10) &  --- &$2.1_{-0.3}^{+0.2}$ & $\bm{-3.7_{-1.6}^{+1.7}}$ &  --- &$0.9_{-0.4}^{+0.6}$ & $-3.7_{-1.8}^{+1.8}$ \\
$c_1$ ({\kms}) & Uniform(-10, 10) &  --- & --- &$\bm{4.0_{-1.1}^{+1.1}}$ &  --- & --- &$3.9_{-1.4}^{+1.4}$ \\
\hline
$\Delta$BIC & --- & 74.0 &  9.2 &  \textbf{0.0} &  1.5 &  3.4 &  3.0  \\
$\Delta$AIC & --- & 73.0 &  8.5 &  \textbf{0.0} &  0.7 &  3.4 &  4.3  \\
\hline
\textbf{Derived} \\
$\psi$ ($^\circ$) & --- & $89.7_{-2.0}^{+2.1}$ &  $89.5_{-2.7}^{+2.8}$ &  $\bm{89.2_{-2.8}^{+3.0}}$ &  $89.6_{-0.3}^{+0.4}$ &  $90.2_{-0.6}^{+0.9}$ &  $90.1_{-3.5}^{+2.1}$ \\
\hline
\enddata
\tablenotetext{}{Values and their uncertainties represent 68\% credible intervals as defined by the 16th, 50th, and 84th percentiles of the sampled posterior. Models with SB rotation do not constrain $\sini$, whereas models with DR prefer large $\alpha$ and small $\sini$ (Fig.~\ref{fig5:cornerplot}). Posteriors in $\psi$ are derived self-consistently using only the posterior samples in $\sini$. That is, they do not incorporate any information about the rotation period. The column with the preferred model is highlighted in bold.\\
$^\ast$Uniform in $\cos\istar$.}
\end{deluxetable*}